\documentclass[12pt,a4paper]{article}
\usepackage{graphicx,rotating}
\usepackage{times}
\usepackage{txfonts}
\usepackage{natbib}

\title{\bf Overview of the BINA Activities}
\author{Santosh Joshi$^1$\thanks{Indian PI of BINA; e-mail: santosh@aries.res.in} \,and  Peter De Cat$^2$\thanks{Belgian PI of BINA; e-mail: Peter.DeCat@oma.be} 
\\
\vspace{1cm}
\\
\normalsize $^1$ Aryabhatta Research Institute of Observational Sciences (ARIES), Manora Peak,\\
\normalsize Nainital-263002, India
\\
\normalsize $^2$ Royal Observatory of Belgium (ROB), Ringlaan 3, 1180 Brussels, Belgium
\\
}
\date{\mbox{}}
\begin{document}
\maketitle
\setcounter{page}{1001}
\pagestyle{plain}
    \makeatletter
    \renewcommand*{\pagenumbering}[1]{%
       \gdef\thepage{\csname @#1\endcsname\c@page}%
    }
    \makeatother
\pagenumbering{arabic}

%
% WE REDEFINE THE plain LaTeX PAGESTYLE !!! 
% THIS PAGESTYLE WILL BE USED FOR THE FIRST PAGE ONLY !
% Please do not change the following lines
%
\def\bull{\vrule height .9ex width .8ex depth -.1ex}
\makeatletter
\def\ps@plain{\let\@mkboth\gobbletwo
\def\@oddhead{}\def\@oddfoot{\hfil\scriptsize\bull\quad
"2nd Belgo-Indian Network for Astronomy \& astrophysics (BINA) workshop'', held in Brussels (Belgium), 9-12 October 2018 \quad\bull}%
\def\@evenhead{}\let\@evenfoot\@oddfoot}
\makeatother
%
% AND DEFINE OUR MACROS FOR THE REFERENCE LIST
% I.E \beginrefer \refer and \endrefer
%
\def\beginrefer{\section*{References}%
\begin{quotation}\mbox{}\par}
\def\refer#1\par{{\setlength{\parindent}{-\leftmargin}\indent#1\par}}
\def\endrefer{\end{quotation}}
%
% BEGIN THE ABSTRACT WITH \noindent\small, ENCLOSE IT IN A GROUP
% AND BOLDFACE THE TITLE.
%
{\noindent\small{\bf Abstract:} 
Here, we summarize the ongoing activities of an international bilateral research project entitled ``Belgo-Indian Network for Astronomy and astrophysics (BINA)'' running jointly since 2014 by the astronomers of various Indian and Belgian institutions. The network activities are being financially supported by the Department of Science and Technology (DST; Government of India) and the Belgian Federal Science Policy Office (BELSPO; Government of Belgium). The structure and mandate of the BINA network are presented. The observational facilities being used to achieve the goal of the project are delineated. The overview of the activities and future perspective in the light of upcoming observational facilities are also highlighted.}
\vspace{0.5cm}\\
% SPECIFY UP TO 5 KEYWORDS SEPARATED BY ' -- '
{\noindent\small{\bf Keywords:} Project : BINA--International bilateral collaboration : -- Area : Astronomy and Astrophysics.}
%
% NOW COMES THE MAIN BODY OF THE ARTICLE
%
%=======
\section{Introduction}
%=======

The Department of Science and Technology (DST), Government of India, is coordinating various collaborative scientific projects with different countries to exchange knowledge in the area of basic sciences and astronomy is one of them. It encourages Indian astronomers to build up joint ventures with the astronomers of other nations to address fundamental questions of our universe such as the mystery of dark energy, extra terrestrial life, etc. 
At ARIES, asteroseismology -- a branch of stellar astrophysics where one can look into the interior of stars by studying the seismic waves at their surface -- is one of the front-line research area where bilateral international projects are being implemented since the last two decades.
The output of these projects have been summarized in the form of research articles published by Ashoka et al. (2000), Balona et al. (2013, 2016), Girish et al. (2001), Joshi et al. (2003, 2006, 2009, 2010, 2012a, 2012b, 2014, 2015, 2016, 2017) and Martinez et al. (1999, 2001). 
One of the major achievement of one of such bilateral project was the development of a {\it Three Channel Fast Photometer} for the 1.04-m Sampurnanand telescope of ARIES for the ``Nainital-Cape Survey'' project (Ashoka et al. 2001).      

\begin{sidewaysfigure}
%\begin{minipage}{15cm}	
\centering
\includegraphics{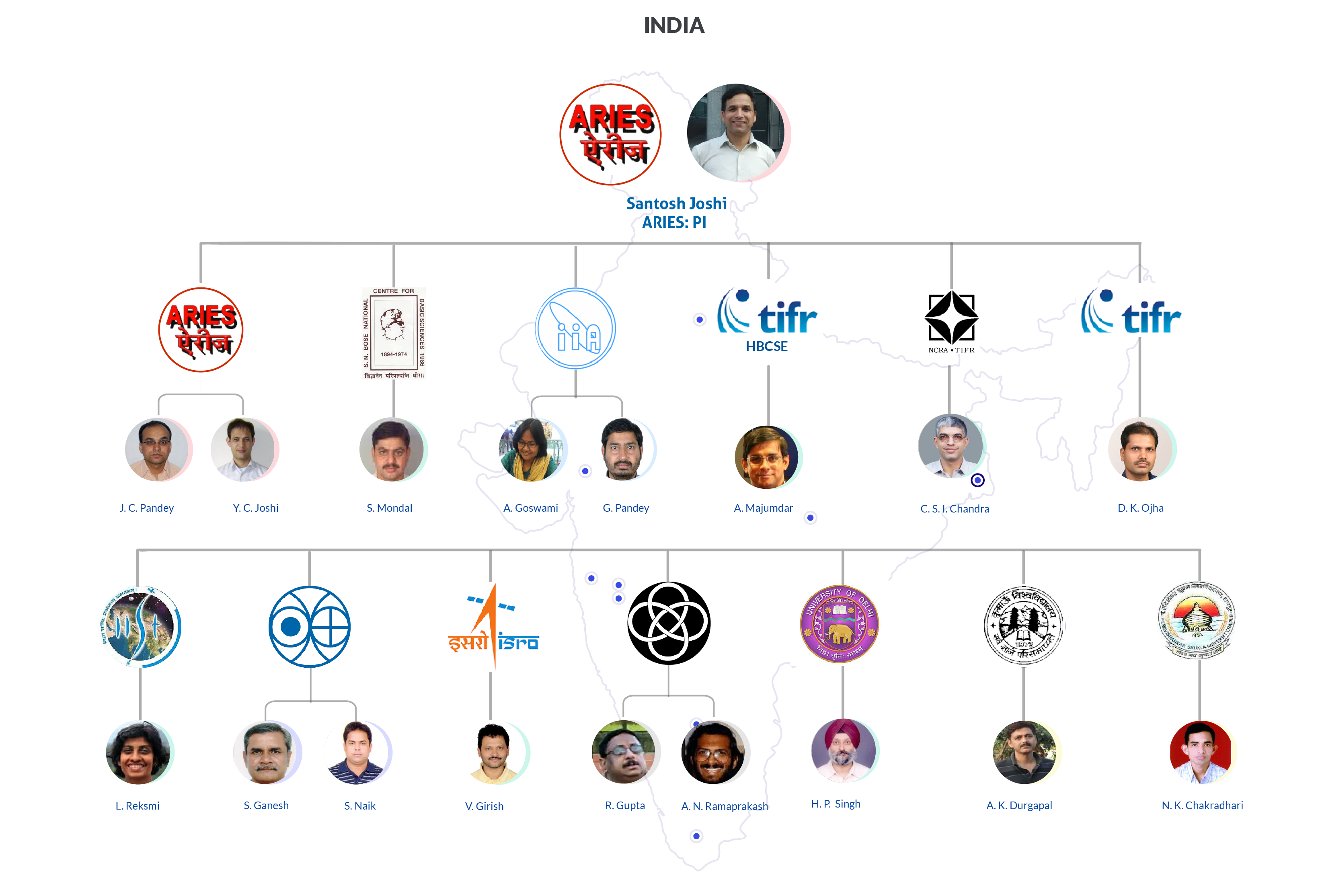}
\caption{The key BINA members from India. The blue dots in the background on the map of India indicate the location of the 13 participating institutes in the different parts of the country.} 
\label{bina_indian}
%\end{minipage}
\end{sidewaysfigure}

The largest observing facility of India in the optical band is the 3.6-m Devasthal Optical Telescope (DOT). It was constructed and commissioned by the company ``Advanced Mechanical and Optical System'' (AMOS; Li\`ege, Belgium). 
This telescope was installed at Devasthal which has been recognized as one of the best observing site based on the extensive site characterization performed between 1980 and 2001 (Stalin et al. 2001). 
This observing facility was technically activated from Brussels on 30 March 2016 jointly by the Indian and Belgian prime ministers and eventually opened for the astronomical community in April 2017. 
The major financial support for this project originated from DST while a share of 2 millions Euro was contributed by BELSPO.
In return, a guaranteed part of 7$\%$ of the available observing telescope time is reserved for the Belgian astronomers to carry out independent research in the various areas of astronomy and astrophysics. 
The 4-m Liquid Mirror Telescope (ILMT) is another project that is in the installation process at Devasthal observatory where the first light is expected to be seen by end of 2019. 
In both the 4-m class telescope projects, Belgian engineers and astronomers are working together for the smooth operation of these observing facilities. 
Taking into account the mutual interest of Belgian and Indian astronomical communities, both DST and BELSPO sanctioned a bilateral project ``Belgo-India Network on Astronomy and Astrophysics (BINA)'' for which these funding agencies have made a huge financial investment. 
Within this network, the Belgian and Indian astronomers are expected to work together for the scientific exploitation of data gathered for solar system, galactic and extra-galactic celestial objects, for the development of new state-of-art instruments, and for the preparation of joint observing proposals to obtain high quality data from these 4-m class telescopes (De Cat et al. 2018). 

\begin{sidewaysfigure}
%\begin{minipage}{15cm}	
\centering
\includegraphics{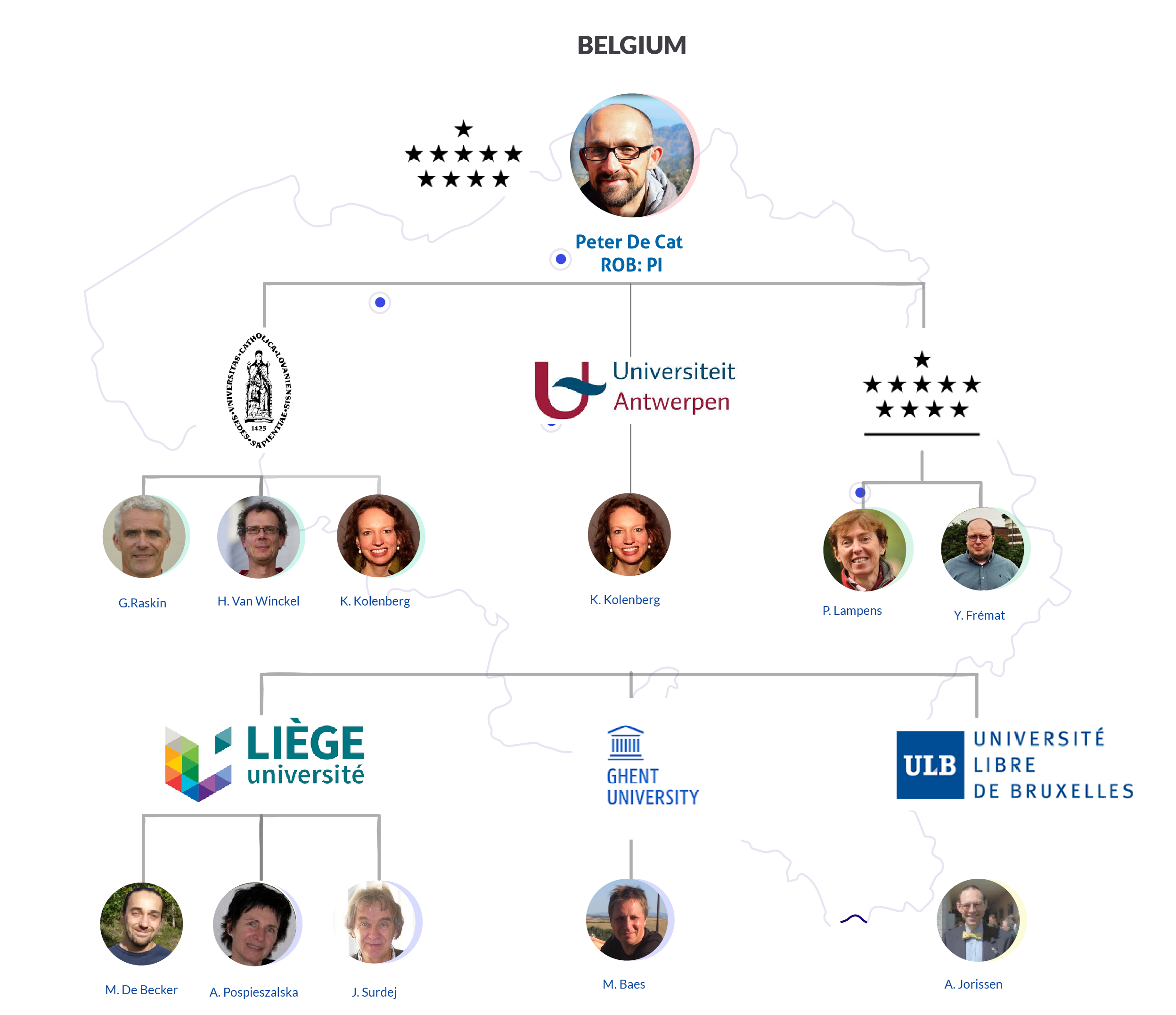}
%%% \caption{The key BINA members from Belgium. The blue dots in the  background of  Belgium map is the location of 5 participating institutes.} 
\caption{Idem as Fig.\,\ref{bina_indian} but for the key BINA members of the 6 participating Belgian institutes.} 
\label{bina_belgian}
%\end{minipage}
\end{sidewaysfigure}

%=======
\section{The Belgo-Indian Network for Astronomy and Astrophysics}
%=======

The backbone of the BINA network are the principal investigators from India and Belgium. The ingredients of this network are described briefly in the following subsections.  

%----------
\subsection{Principal Investigators and Partner Institutions}
%----------

%%% PDC: the number of institutes in the text and on the figures should be the same.
BINA unites astronomers from 13 Indian and 6 Belgian institutions, known as the BINA members. 

At the Indian side, the participating astronomical organizations are funded by 
the DST (\#3: Aryabhatta Research Institute of Observational Sciences (ARIES; Nainital), Indian Institute of Astrophysics (IIA; Bengaluru) and S. N. Bose National Center for Basic Sciences (SNBNCBS; Kolkata)), 
the Department of Space (\#3: Indian Space Research Organization (ISRO; Bengaluru), Physical Research Laboratory (PRL; Ahmedabad), and Indian Institute of Science and Technology (IIST; Thiruvananthapuram)),
the Department of Atomic Energy (\#3: Tata Institute of Fundamental Research (TIFR; Mumbai), National Center for Radio Astronomy (NCRA; Pune), and Homi Bhabha Centre for Science Education (HBCSE; Mumbai)), and 
the University Grant Commission (\#4: Inter University Center for Astronomy and Astrophysics (IUCAA; Pune), Delhi University (DU; Delhi), Kumaun University (KU; Nainital), and Pt. Ravi Shankar University (RSU; Raipur)).
The flow chart of the key Indian BINA members and their respective affiliation are shown in Fig.\,\ref{bina_indian}. 

At the Belgian side, the participating astronomical organizations are one federal scientific institute (\#1: Royal Observatory of Belgium (ROB; Brussels)),
French-speaking universities (\#2: Universit\'e de Li\`ege (ULi\`ege; Li\`ege) and Universit\'e Libre de Bruxelles (ULB; Brussels)), and
Dutch-speaking universities (\#3: Katholieke Universiteit Leuven (KU Leuven; Leuven), Universiteit Gent (UGent; Ghent), and Universiteit Antwerpen (UAntwerpen; Antwerp)). Fig.\,\ref{bina_belgian} shows the scientists representing the various Belgian organizations which are BINA members. 

ARIES and ROB are the leading institutes from where the BINA activities are being governed.   
This is the first-of-its-kind bilateral project where such large numbers of investigators from two countries are collaborating to carry out the fundamental science in the area of astronomy and astrophysics.

%----------
\subsection{Telescopes of Interest}
%----------

The scientific projects within BINA involve observations of celestial objects emitting light with wavelengths across the electro-magnetic spectrum. 
To accomplish the goals, extensive observations in different wavelength ranges can be done using the space and ground based observational facilities available within network. 
Apart from the 3.6-m DOT and 4-m ILMT there are many other telescopes of interest those are accessible for BINA members through the network. 

There are a lot of observing facilities at the Indian territory.
The Indian BINA institutes are operating four optical telescopes (2.34-m Vainu Bappu Telescope at Kavalur, 2.01-m Himalayan Chandra Telescope at Leh, 2.0-m IUCAA telescope at Girawali and 1.3-m telescope at Devesthal), 1.2-m near infrared (NIR) telescope at Mount Abu Observatory, Rajasthan, and a radio interferometer the Giant Metrewave Radio Telescope (GMRT) at Pune.
The first Indian space observatory `Astrosat', operational in the ultraviolet (UV) and X-rays is also available to the BINA members. 
Moreover, Indian astronomers have access to the 6.0-m optical telescope of the Special Astrophysical Observatory (SAO) in Russia under collaborative programmes sponsored by the Russian Foundation for Basic Research (RFBR) and DST. The 10-m optical telescopes of the South African Astronomical Observatory located at Sutherland is also approachable through the SALT consortium where India is an international partner.

\begin{figure}
%\begin{minipage}{15cm}	
\centering
\includegraphics[width=15cm]{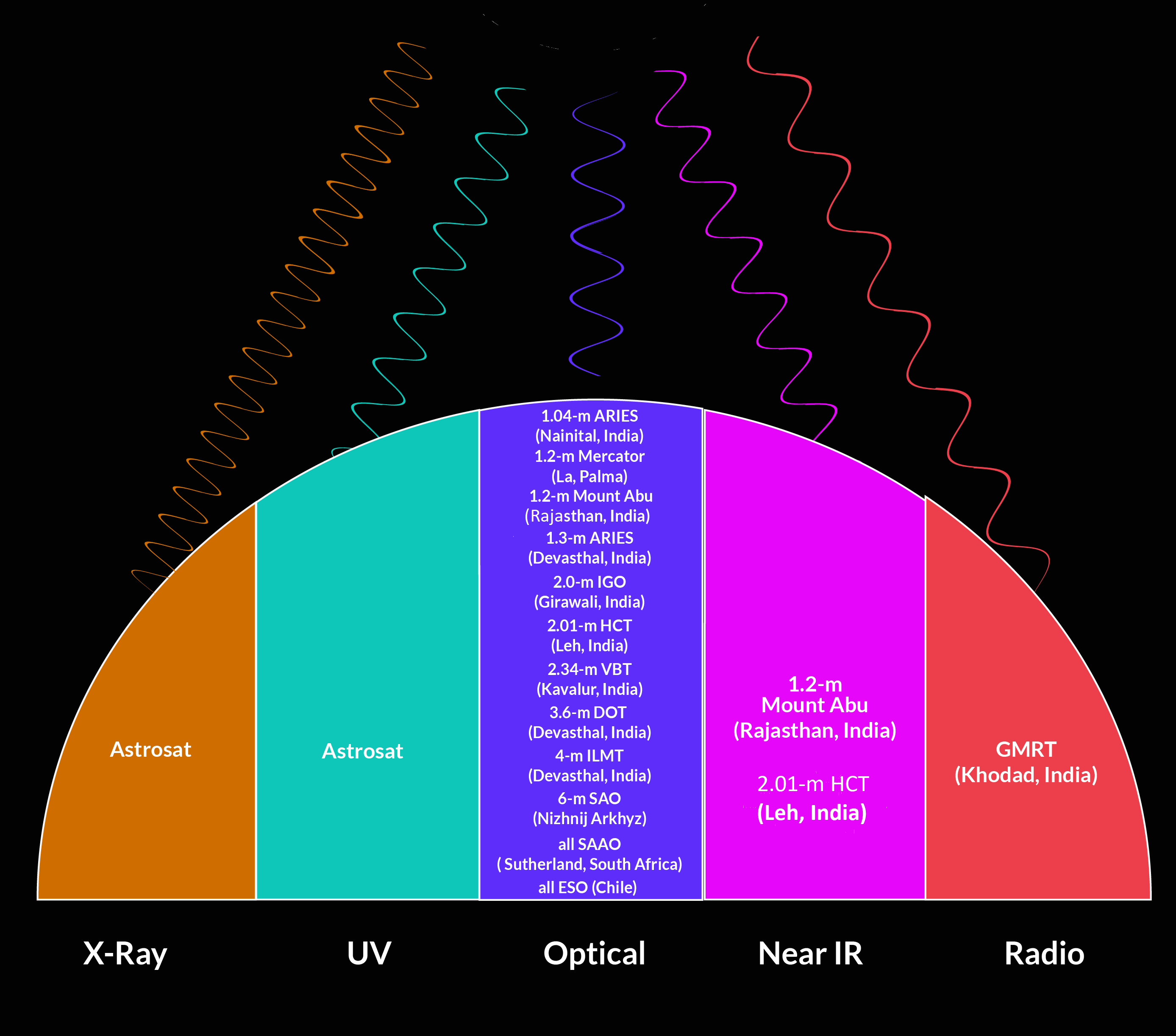}
%\vspace{-3.0cm}
\caption{The ground and spaced based observation facilities being used by the BINA members to address the scientific rationale proposed for the BINA network.} 
\label{telescopes}
%\end{minipage}
\end{figure}

Belgium has also additional observational facilities to offer.
An optical 1.2-m Mercator telescope located at the Canary-island of La Palma (Spain), is accessible via the HERMES consortium lead and operated by KU Leuven. 
%%% PDC: Note that ESO is not restricted to La Silla only because you also have Paranal (with e.g. VLTs), Chajnantor (with ALMA & APEX) and (in the future) Cerro Armazones (with ELT). 
Since Belgium is a member of ESO, hence Belgian astronomers can apply for observing time on all ESO telescopes located in Chile.

This clearly illustrates that BINA is not restricted to India and Belgium, but it is an expanding collaboration initiated under the impulse of Indian and Belgian astronomers. 
The telescopes equipped with back-end instruments currently being used and/or foreseen to be used within BINA to investigate celestial bodies are shown in Fig.\,\ref{telescopes}. 
A new optical telescope of  2.5-m aperture is being built by AMOS and will be installed at the mountain of Mount Abu, Rajasthan by 2020. The proposed back-end instruments for this telescope are a high-resolution spectrograph (PARAS2), an infrared imager, FOSC and a spectrometer and polarimeter (NISP). This telescope would be used for a wide variety of scientific programs such as understanding the star formation process, study of exo-planets and investigating the galactic and extra-galactic sources as well as solar system bodies.

%----------
\subsection{Funding}
%----------

%%% In the original BINA proposal the major financial support was sought for the bilateral conference, exchange visits etc.  The DST and BELSPO sanctioned Rs. 19,84,600/- and  29,085/- Euros respectively to ARIES and ROB those are the leading institutions of India and Belgium, respectively. The given financial support was for a period of 3 years beginning from 2014 and extendable for one more year based on the progress of the project activities. The statistical distribution of budget as received for the execution of BINA activities from the funding agencies  are shown in Fig. \ref{budget} using a pie chart.

\begin{figure}
%\begin{minipage}{15cm}	
\centering
\includegraphics[width=15cm]{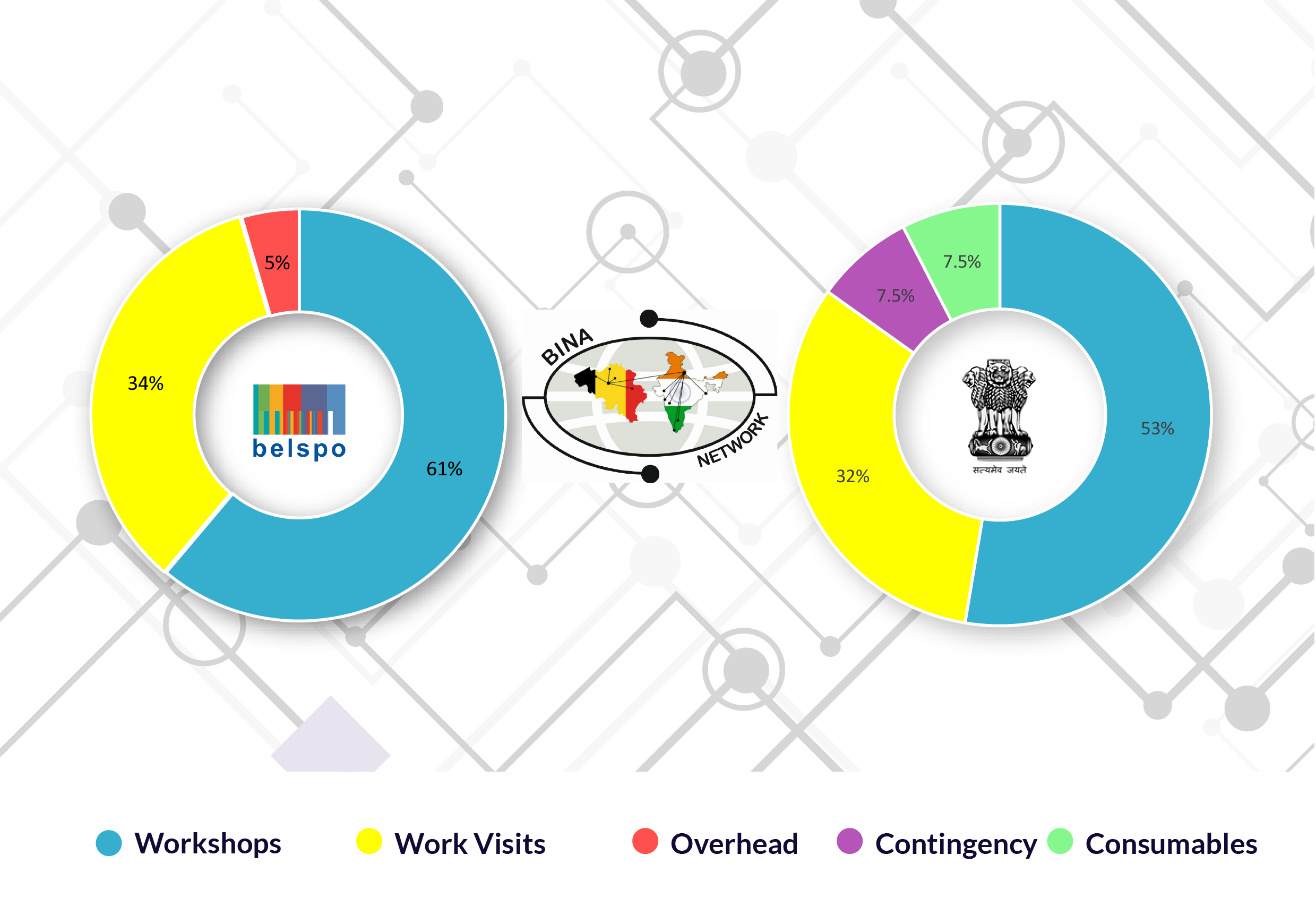}
\caption{
A pie chart showing the budgetary scheme for the various BINA activities as sanctioned by DST and BELSPO. 
%The major part of the budget is allocated to the project seminar that show the importance of face-to-face interaction of the scientists from both the countries.
} 
\label{budget}
%\end{minipage}
\end{figure}

For the first phase of BINA (BINA-1), the DST and BELSPO sanctioned 19,84,600/- INR and 29,085/- EURO to ARIES and ROB, respectively. 
The statistical distribution of the budget as received for the execution of BINA networking activities in both countries is shown in Fig.\,\ref{budget} using a pie chart.
BINA officially started in Belgium on 15/12/2014 and in India on 05/05/2016. BINA-1 ends on 15/12/2018 and 05/05/2019 in Belgium and India, respectively. However, the project is already extended for another three years in both countries (BINA-2; see Section\,\ref{BinaFuture}).

%----------
\subsection{Network Activities}
%----------

%%% In the following subsections we have outlined the major activities those we have been conducting in last couple of years to meet the objective of the project and the same are depicted in Fig. \ref{activity}.

\begin{sidewaysfigure}
%\begin{minipage}{15cm}	
\centering
\includegraphics[width=27cm]{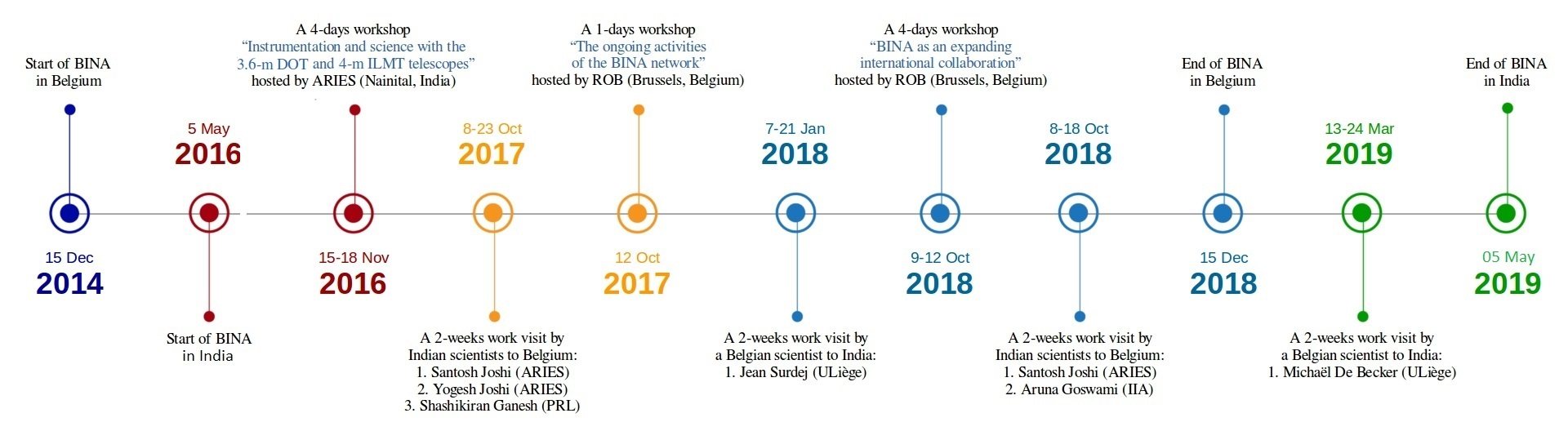}
\caption{The BINA network activities organized in the first phase of the project are shown in chronological order. The activities of different years are shown by different colors. The workshops and work visits are given above and below the time line, respectively.} 
\label{activity}
%\end{minipage}
\end{sidewaysfigure}

The BINA funding foresees financial supports for network activities including the organization of workshops (Section\,\ref{BinaWorkshop}) and work visits (Section\,\ref{BinaWorkVisit}), allowing face-to-face interactions of colleagues from both countries to establish mutual collaborations in order to work on joint science projects (Section\,\ref{BinaScienceProject}) and on the development of new back-end instruments for the new telescopes (Section\,\ref{BinaInstrumentDevelopment}). 
In this section, the major network activities within BINA-1 are described and shown in chronological order in Fig.\,\ref{activity}.

%.............
\subsubsection{Workshops} \label{BinaWorkshop}
%.............

%The largest part of the BINA funding is used to organise workshops of typically four days.

%\paragraph{First BINA workshop}
{\it First BINA workshop}:
%The network activities of BINA officially started with the first BINA workshop ({\tt https://aries.res.in/{\raise.17ex\hbox{$\scriptstyle\sim$}}bina/}).schematic
The first BINA network activity ({\tt https://aries.res.in/{\raise.17ex\hbox{$\scriptstyle\sim$}}bina/}) was a 4-days workshop hosted by ARIES in Nainital (India) and took place from 15 to 18 November 2016. 
The theme of this workshop was ``Instrumentation and Science with the 3.6-m DOT and 4-m ILMT''. 
There were 107 participants from 8 different countries, including \#88 from India and \#10 from Belgium.
%The main reason behind organizing this workshop was to have a face-to-face interaction of Belgian astronomers with the large Indian astronomical community for strengthening the on-going collaborations, starting the new ones and to be familiar with the existing instruments and have discussion on the future requirement for the 3.6-m DOT and other telescopes of interest. 
This was the ideal opportunity for the Belgian participants to get the first face-to-face contact with members of the large Indian astronomical community.
During the workshop, an overview of the existing instruments of the Indo-Belgian and other telescopes of interest was given and the requirements for the next-generation instrumentation were discussed. 
Members of the BINA institutes had the opportunity to introduce their scientific interests to the participants.
This allowed to strengthen ongoing collaborations and to start new ones.
The proceedings of this conference were published in the peer-reviewed journal {\it Bulletin de la Soci\'et\'e Royale des Sciences de  Li\`ege} and are freely available online\footnote{The articles of the proceedings of the first BINA workshop can be downloaded from {\tt https://popups.uliege.be/0037-9565/index.php?id=7434}}.
The highlights of this workshop are summarized by De Cat et al. (2018). 

%\paragraph{Second BINA Workshop}
{\it Second BINA Workshop}:
The ROB hosted the second BINA workshop in Brussels (Belgium) from 9 to 12 October 2018 ({\tt http://aa.oma.be/18bina}). 
A total of 65 scientists originating from 7 different countries participated to the workshop, including \#37 from India and \#23 from Belgium.
The aim of this workshop was to give a full overview of ``BINA as an expanding international collaboration''.
This included an update of the available and upcoming astronomical instruments accessible through BINA during the first day.
The rest of the time was devoted to discuss the ongoing and future science projects based on data gathered with the 3.6-m DOT and other telescopes of interest.
A more detailed description of this workshop is given by De Cat et al. (2019).

%https://events.oma.be/indico/event/32/

%.............
\subsubsection{Work visits} \label{BinaWorkVisit}
%.............

BINA provides financial supports to the members of the Indian and Belgian BINA institutes to visit each other typically for a period of one to two weeks to stimulate face-to-face discussions and review the progress of ongoing science projects and the development of new instrumentation.
Such work visits are particularly useful to increase the exchange of scientific knowledge, to explore the future prospective of the collaborative projects, and to discuss the outcome of the scientific results. 
Work visits can also be combined with conferences or observation runs.
By joining forces and exchanging experience about observing, data reduction, and analysis techniques, the BINA members will be able to obtain more insight into the data that are being collected for projects of joint scientific interest.
Moreover, the students associated with BINA institutes can broaden their views and hence can be attracted to a scientific career in the area of astronomy and astrophysics.
Several working visits have been sponsored during the first phase of the BINA project.

%\paragraph{From India to Belgium}
{\it From India to Belgium}:
The first work visit took place in October 2017 when three Indian colleagues came to Belgium for two weeks. Santosh Joshi (ARIES), Yogesh Joshi (ARIES), and Sashikiran Ganesh (PRL) stayed at the ROB to initiate  new joint observational projects for the 3.6-m DOT and 2.01-m HCT. A 1-day workshop was organized during their visit where 17 BINA members highlighted the ongoing activities of the BINA. Also an update about the design of the high-resolution spectrograph (HRS) for the 3.6-m DOT was given. Nearly all the Belgian BINA institutes were presented during this meeting. 
In October 2018, Santosh Joshi (ARIES) and Aruna Goswami (IIA) stayed an extra week in Belgium after the second BINA workshop to discuss about the progress and future steps in the analysis of the data gathered for a joint project with colleagues from the ROB. Aruna Goswami also visited to the colleagues at the ULB for scientific discussion.

On the other hand, a mobility programme is being coordinated by Micha\"el De Becker at ULi\`ege to allow PhD students from India to have a stay of three months in Li\`ege in the framework of scientific collaborations involving BINA institutes. The mobility grants are funded by the Erasmus+ programme of the European Union. To date, three PhD students (from ARIES, IUCAA, and IIA) have already been benefited from such a mobility grants in 2018, and other grants are in the process to be distributed to ARIES students in 2019. This fruitful mobility programme is intended to be pursued in the forthcoming years.

%\paragraph{From Belgium to India}
{\it From Belgium to India}:
In January 2018, Jean Surdej (ULi\`ege) went to the Devasthal Observatory for two weeks to perform installation activities for the 4-m ILMT. In March 2019, Micha\"el De Becker (ULi\`ege) visited  India for scientific discussions with scientists of IIST, Thiruvananthapuram and NCRA Pune. He completed his work visit by attending an international  conference on ``The Meter Wavelength Sky-II'' from 17 to 24 March 2019 at NCRA in Pune. 

\begin{figure}
%\begin{minipage}{15cm}	
\centering
\includegraphics[width=19cm]{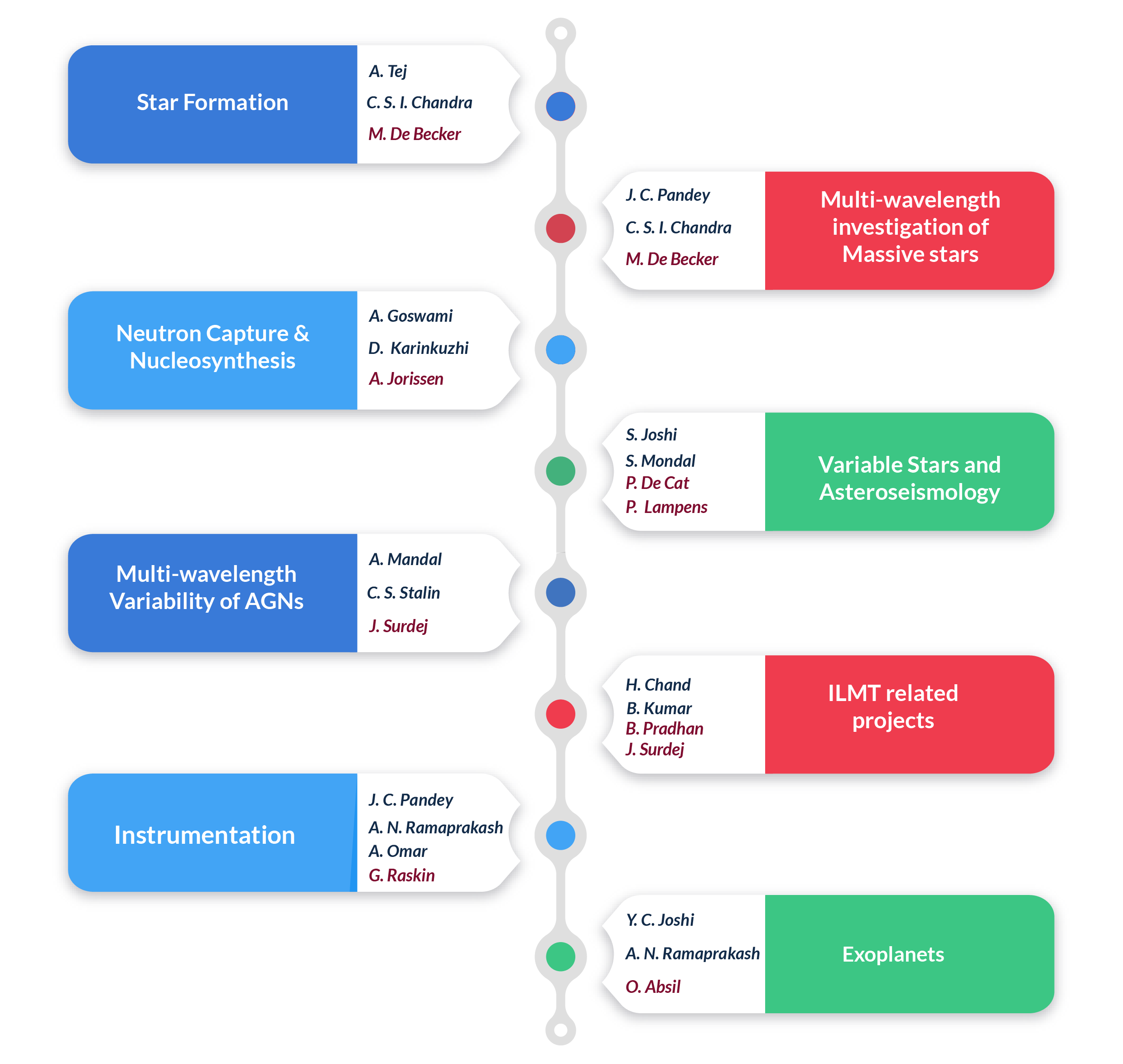}
\caption{
A schematic overview of the main research areas for which Indian and Belgian BINA members are collaborating to each other. The names of the key Indian and Belgian investigators are highlighted in black and red colors, respectively.
}
\label{science}
%\end{minipage}
\end{figure}

%----------
\subsection{Science Projects} \label{BinaScienceProject}
%----------

The BINA members from both countries are actively involved in different areas of astronomy and astrophysics through study of different kinds of astronomical objects at wavelengths ranging from $\gamma$-rays to radio. 
Indeed,  complete understanding of the astronomical/studied objects can only be obtained by using multi-band imaging and high-resolution spectrometry/astrometry. 
The scientific and instrumentation areas where the opportunities exist for joint collaborations  are depicted in Fig.\,\ref{science}.

%----------
\subsection{Instrument Development} \label{BinaInstrumentDevelopment}
%----------

In observational astronomy, state-of-the art instruments are highly essential to carry out front line research. The 4-m class telescopes  are efficient in collecting photons from the sky. They are located in a new observatory in Devasthal at an altitude of 2450 meters in the Kumaun region of the Himalayas towards north of India. The astronomical conditions of this site are excellent and the location is ideal to fill a gap in longitude that is currently not well-covered by other telescopes. If these new telescopes are equipped with a carefully chosen arsenal of instruments, then the Devasthal observatory would be recognized as one of the world-leading astronomical observing station for the study of cosmic events. The BINA members are combining their forces to optimize the design for new instruments for the new telescopes at the Indian observatories.

%=======
\section{Future Prospects} \label{BinaFuture}
%=======

Based on the progress and outcome of the BINA-1 activities, the DST and BELSPO have already approved the second phase of the BINA project. BINA-2 can cover network activities up to 2021. There is financial support for a total of ten 2-weeks work visits (five in each direction) and the organization of the third BINA workshop with the aim to demonstrate the potential use of the telescopes of interest. The Indian and Belgian partners (and their international collaborators) will get the opportunity to present the scientific results obtained for projects that make use of the observational facilities of mutual interest. The need for the (further) development of new instruments for the 3.6-m DOT will be also evaluated. The third BINA workshop is planed to be organized in India and is currently scheduled in the second semester of 2021.

%=======
\section{Conclusions}
%=======

Astronomy and astrophysics is one of the areas where researchers from India and Belgium are working together to enhance their scientific capability and strengthen the technology being used in the instrumentation. 
BINA attempts to optimize the use of the new telescopes installed at the Devasthal observatory and to equip them with top-quality instruments.
Hence, the BINA project has a scientific, technical and educational values.
In order to measure and fully understand the nature of radiation from celestial objects, the high-resolution spectroscopy and seeing limited imaging in the visible and near infrared bands are indispensable.
Observations with the 3.6-m DOT are therefore of great interest provided that a good choice of instruments is offered.
With the 4-m ILMT at Devasthal, we expect to discover many transient sources through the deep imaging of a portion of sky.  
The other telescopes of interest can be used to monitor the interesting sources  in photometric and spectroscopic mode to allow their follow-up studies.

By exchanging and reinforcing our expertise and observational capabilities in multi-wavelengths astronomy, we will be able to address many fundamental questions related to the cosmos such as the origin of the galaxies, the life-cycles and the internal structure of stars, the search and study of new exo-planets akin to Earth, probing the powerful and elusive black holes, search for the first stars and galaxies in the universe, use of gravitational lenses to probe the structure of the most distant objects as well as to test the cosmological models, etc. 
Now that the main scientific areas of mutual interest have been identified (Fig.\,\ref{science}), the focus can move more towards the joint supervision, training and guidance of PhD students associated to the BINA partner institutes. The methodologies can be further developed by students of the new generation. In this way, we can ensure a long and successful future for our joint venture!

%
% USE A SECTION WITHOUT NUMBER FOR THE ACKNOWLEDGEMENTS
%
\section*{Acknowledgements}
The authors highly acknowledge the tremendous support received from BINA members of India and Belgium. The work presented in these proceedings is supported by the Belgo-Indian Network for Astronomy \& Astrophysics (BINA), approved by the International Division, Department of Science and Technology (DST, Govt. of India; DST/INT/Belg/P-02) and the Belgian Federal Science Policy Office (BELSPO, Govt. of Belgium; BL/11/IN07). SJ appreciate effort of Ms. Shilpi Jain from IIT, Roorkee for preparing the impressive figures and dedicates this article to his younger brother Navin ``Guddu'' who passed away about a year ago.
%
% BEGIN THE REFERENCE LIST WITH \beginrefer
% USE \refer BEFORE THE REFERENCES AND BEGIN A NEW PARAGRAPH AFTER THE 
% REFERENCE !
% DO NOT FORGET TO END THE LIST WITH \endrefer
% 
%
% INSTRUCTIONS FOR BIBLIOGRAPHY:
% ==============================
% - DON'T USE THE & SYMBOL
% - USE INITIALS FOR FIRST AND MIDDLE NAMES, AND SPECIFY FULL FAMILY NAME (see examples below)
% - NO COMMA BETWEEN NAME AND INITIALS
% - USE COMMA BETWEEN DIFFERENT AUTHORS NAMES
% - NO COMMA AFTER THE LAST AUTHOR NAME
% - FOR LONG AUTHOR LISTS, SPECIFY THE FIRST 3 AUTHORS FOLLOWED BY 'et al.', WITH NO COMMA BEFORE AND AFTER 'et al.'
% - INSERT A BLANK SPACE BETWEEN MULTIPLE INITIALS
% - USE STANDARD JOURNAL ACRONYMS FREQUENTLY USED IN MAIN ASTROPHYSICS JOURNAL
% - SORT REFERENCES BY ALPHABETICAL ORDER OF FIRST AUTHOR NAMES

\footnotesize
\beginrefer

\refer Ashoka B. N., Seetha S., Raj E. et al. 2000, Bulletin of the Astronomical Society of India, 28, 251 %*

%%% \refer Ashoka B. N., Kumar Babu V. C., Seetha S., Girish V. et al 2001, JApA, 22, 131
\refer Ashoka B. N., Kumar Babu V. C., Seetha S. et al 2001, JApA, 22, 131 %*
 
%%% \refer Balona L.A., Catanzaro G., Crause L., Cunha M.S. et al. 2013, MNRAS, 432, 2808
\refer Balona L. A., Catanzaro G., Crause L. et al. 2013, MNRAS, 432, 2808 %*

%%% \refer Balona L.A., Engelbrecht C.A. et al. 2016, MNRAS, 460, 1318 
\refer Balona L. A., Engelbrecht C. A., Joshi Y. C. et al. 2016, MNRAS, 460, 1318 %* 

%\refer Chowdhury S., Joshi S., Engelbrecht C. A. et al. 2018, Ap\&SS, 363, 260

\refer De Cat P., Surdej J., Omar A., De Becker M., Joshi S. 2018, BSRSL, 87, 1

\refer De Cat P., Lampens P., De Becker M. et al. 2019, BSRSL, 88, 1

\refer Girish V., Seetha S., Martinez P. et al. 2001, A\&A, 380, 142 %*
 
%%% \refer Joshi S., Joshi, Y. C. 2015, JA\&A, 36, 33
\refer Joshi S., Joshi Y. C. 2015, JA\&A, 36, 33 %*
 
\refer Joshi S., Girish V., Sagar R. et al. 2003, MNRAS, 344, 431 %*

%%% \refer Joshi S. et al. 2006, A\&A, 455, 303
\refer Joshi S., Mary D. L., Martinez P. et al. 2006, A\&A, 455, 303 %*
 
\refer Joshi S., Mary D. L., Chakradhari N. K., Tinari S. K., Billaud C. 2009, A\&A, 507, 1763 %* 
 
\refer Joshi S., Ryabchikova T., Kochukhov O. et al. 2010, MNRAS, 401, 1299 %*
 
\refer Joshi S., Semenko E., Martinez P. et al. 2012a, MNRAS, 424, 2002 %*
 
%%% \refer Joshi S., Martinez P., Chowdhury S., et al. 2016, A\&A, 590, A116
\refer Joshi S., Martinez P., Chowdhury S. et al. 2016, A\&A, 590, A116 %*

%%% \refer Joshi S., Semenko E., Moiseeva A., et al. 2017, MNRAS, 467, 633 
\refer Joshi S., Semenko E., Moiseeva A. et al. 2017, MNRAS, 467, 633 %*
 
%%%
\refer Joshi Y. C., Joshi S., Kumar B., Mondal S., Balona L. A. 2012b, MNRAS, 419, 2379 %*
 
\refer Joshi Y. C., Balona L. A., Joshi S., Kumar B. 2014, MNRAS, 437, 804 %*
 
%%% \refer Martinez P., Kurtz D. W., 1994, MNRAS, 271, 129
\refer Martinez P., Kurtz D. W. 1994, MNRAS, 271, 129
 
\refer Martinez P., Kurtz D. W., Ashoka B. N. et al. 1999, MNRAS, 309, 871 %*

\refer Martinez P., Kurtz D. W., Ashoka B. N. et al. 2001, A\&A, 371, 1048 %*   
  
\refer Stalin C. S., Sagar R., Pant P. et al. 2001, BASI, 29, 39
  
\endrefer           

\end{document}